\documentclass[onecolumn,showpacs,showkeys,preprintnumbers,amsmath,amssymb]{revtex4}

\usepackage{graphicx}
\usepackage{dcolumn}
\usepackage{bm}
\usepackage{graphics}

\begin{document}
\draft
\title{Enhanced transmission versus localization of a light pulse by a
subwavelength\\
metal slit: Can the pulse have both characteristics?}
\author{S. V.  Kukhlevsky$^a$, M. Mechler$^b$, L. Csapo$^c$, K. Janssens$^d$, O. Samek$^e$}
\address{$^a$Institute of Physics, University of Pecs, Ifjusag u. 6, Pecs 7624,
Hungary\\
$^b$South-Trans-Danubian Cooperative Research Centre,
University of Pecs, Ifjusag u. 6, Pecs 7624, Hungary\\
$^c$Institute ofMathematics and Information, University of Pecs,
Ifjusag u. 6, Pecs 7624, Hungary\\
$^d$Department of Chemistry, University of Antwerp,
Universiteitsplein 1, B-2610 Antwerp, Belgium\\
$^e$Institute of Spectrochemistry and Applied Spectroscopy,
Bunsen-Kirchhoff-Str. 11, D-44139 Dortmund, Germany}
\begin{abstract}
The existence of resonant enhanced transmission and collimation of
light waves by subwavelength slits in metal films [for example,
see T.W. Ebbesen et al., Nature (London) 391, 667 (1998) and H.J.
Lezec et al., Science, 297, 820 (2002)] leads to the basic
question: Can a light be enhanced and simultaneously localized in
space and time by a subwavelength slit? To address this question,
the spatial distribution of the energy flux of an ultrashort
(femtosecond) wave-packet diffracted by a subwavelength
(nanometer-size) slit was analyzed by using the conventional
approach based on the Neerhoff and Mur solution of Maxwell's
equations. The results show that a light can be enhanced by orders
of magnitude and simultaneously localized in the near-field
diffraction zone at the nm- and fs-scales. Possible applications
in nanophotonics are discussed.
\end{abstract}

\maketitle
\section{Introduction}
In the last decade nanostructured optical elements based on
scattering of light waves by subwavelength-size metal objects,
such as particles and screen holes, have been investigated,
intensively. The most impressive features of the optical elements
are resonant enhancement and spatial localization of optical
fields by the excitation of electron waves in the metal (for an
example see,
e.g.~\cite{Neer,Harr,Betz1,Ash,Lewi,Betz2,Pohl,Ebbe,Leze,Port,Hibb,
Gar1,Gar2,Dog,Li,Stoc,Kuk1,Kuk2,Stav,Bori,Taka,Yang,Dykh,Cao,Alte}).
Recently, some nanostructures namely a single subwavelength slit,
a grating with subwavevelength slits and a subwavelength slit
surrounded by parallel deep and narrow grooves attracted a
particular attention of researchers. The study of resonant
enhanced transmission and collimation of waves in close proximity
to a single subwavelength slit acting as a microscope probe
\cite{Neer,Harr,Betz1} was connected with developing near-field
scanning microwave and optical microscopes with subwavelength
resolution \cite{Ash,Lewi,Betz2,Pohl}. The resonant transmission
of light by a grating with subwavevelength slits and a
subwavelength slit surrounded by grooves is an important effect
for nanophotonics \cite{Ebbe,Leze,Port,Hibb}. The
transmissitivity, on the resonance, can be orders of magnitude
greater than out of the resonance. It was understood that the
enhancement effect has a two-fold origin: First, the field
increases due to a pure geometrical reason, the coupling of
incident plane waves with waveguide mode resonances located in the
slit, and further enhancement arises due to excitation of coupled
surface plasmon polaritons localized on both surfaces of the slit
(grating) \cite{Port,Hibb,Gar1}. A dominant mechanism responsible
for the extraordinary transmission is the resonant excitation of
the waveguide mode in the slit giving a Fabry-Perot like behavior
\cite{Hibb}. In addition to the extraordinary transmission, a
series of parallel grooves surrounding a nanometer-size slit can
produce a micrometer-size beam that spreads to an angle of only
few degrees \cite{Leze}. The light collimation, in this case, is
achieved by the excitation of coupled surface plasmon polaritons
in the grooves \cite{Gar1}. At appropriate conditions, a single
subwavelength slit flanked by a finite array of grooves can act as
a "lens" focusing a light \cite{Gar2}. It should be noted, in this
connection, that the diffractive spreading of a beam can be
reduced also by using a structured aperture or an effective
nanolens formed by self-similar linear chain of metal nanospheres
\cite{Dog,Li}.

New aspects of the problem of resonantly enhanced transmission and
collimation of light are revealed when the nanostructures are
illuminated by an ultra-short (fs) light pulse
\cite{Stoc,Kuk1,Kuk2,Stav,Bori}. For instance, in the study
\cite{Stoc}, the unique possibility of concentrating the energy of
an ultrafast excitation of an "engineered" or a random nanosystems
in a small part of the whole system by means of phase modulation
of the exciting fs-pulse was predicted. The study \cite{Kuk1}
theoretically demonstrated the feasibility of nm-scale
localization and distortion-free transmission of fs visible pulses
by a single metal slit, and further suggested the feasibility of
simultaneous super resolution in space and time of the near-field
scanning optical microscopy (NSOM). The quasi-diffraction-free
optics based on transmission of pulses by a subwavelength
nano-slit has been suggested to extended the operation principle
of a 2-D NSOM to the "not-too-distant" field regime (up to ~0.5
wavelength) \cite{Kuk2}. Some interesting effects namely the pulse
delay and long leaving resonant excitations of electromagnetic
fields in the resonant-transition gratings were recently described
in the studies \cite{Stav,Bori}.

The great interest to resonant enhanced transmission, spatial
localization (collimation) of continuous waves and light pulses by
subwavelength metal slits leads to the basic question: Can a light
be enhanced and simultaneously localized in space and time by a
subwavelength slit? If the field enhancement can be achieved
together with nm-scale spatial and fs-scale temporal
localizations, this could greatly increase a potential of the
nanoslit systems in high-resolution applications, especially in
near-field scanning microscopy and spectroscopy. In the present
article we test whether the resonant enhancement could only be
obtained at the expense of the spatial and temporal broadening of
a light wavepacket. To address this question, the spatial
distribution of the energy flux of an ultrashort (fs) pulse
diffracted by a subwavelength (nanosized) slit in a thick metal
film of perfect conductivity will be analyzed by using the
conventional approach based on the Neerhoff and Mur solution of
Maxwell's equations. In short, we first will describe the
theoretical development of Neerhoff and Mur (Section 2) and the
model will then be used to calculate the spatial distribution of
the energy flux of the transmitted pulse (wavepacket) under
various regimes of the near-field diffraction (Section 3). We will
show that a light can be enhanced by orders of magnitude and
simultaneously localized in the near-field diffraction zone at the
nm- and fs-scales. The implications of the results for
diffraction-unlimited near- and far-field optics will then be
discussed. In Section 4 we summarize results and present
conclusions.

\section{Theoretical background}
An adequate description of transmission a light by a subwavelength
nano-sized slit in a thick metal film requires solution of
Maxwell's equations with complicated boundary conditions.  The
light-slit interaction problem even for a continuous wave can be
solved only by extended two-dimensional $(x,z)$ numerical
computations. The tree-dimensional $(x,z,t)$ character of the
pulse-slit interaction makes the numerical analysis even more
difficult. Let us consider the near-field diffraction of an
ultrashort pulse (wave-packet) by a subwavelength slit in a thick
metal screen of perfect conductivity by using the conventional
approach based on the Neerhoff and Mur solution of Maxwell's
equations. Before presenting a treatment of the problem for a
wave-packet, we briefly describe the Neerhoff and Mur formulation
\cite{Neer,Betz1} for a continuous wave (a Fourier
$\omega$-component of a wave-packet). The transmission of a plane
continuous wave through a slit (waveguide) of width $2a$ in a
perfectly conducting screen of thickness $b$ is considered. The
slit is illuminated by a normally incident plane wave under TM
polarization (magnetic-field vector parallel to the slit), as
shown in Fig. 1. The magnetic field of the wave is assumed to be
time harmonic and constant in the $y$ direction:

\begin{eqnarray}
{\vec{H}}(x,y,z,t)=U(x,z){\exp}(-i\omega{t}){\vec{e}}_y.
\end{eqnarray}
The electric field of the wave is found from the scalar field $U(x,z)$
using Maxwell's equations:
\begin{eqnarray}
{E_x}(x,z,t)=-{\frac{ic}{\omega{\epsilon_1}}}{{\partial}_z}{U(x,z)}{\exp}(-i\omega{t}),
\end{eqnarray}
\begin{eqnarray}
{E_y}(x,z,t)=0.
\end{eqnarray}
\begin{eqnarray}
{E_z}(x,z,t)=-{\frac{ic}{\omega{\epsilon_1}}}{{\partial}_x}{U(x,z)}{\exp}(-i\omega{t}).
\end{eqnarray}
Notice that the restrictions in Eq. 1 reduce the diffraction problem to one involving
a single scalar field $U(x,z)$ in two dimensions. The field is represented by $U_{j}(x,z)$
($j$=1,2,3 in each of the three regions I, II and III). The field satisfies the
Helmholtz equation:
\begin{eqnarray}
({\nabla}^2+k_{j}^2)U_j=0,
\end{eqnarray}
where $j=1,2,3$. In region I, the field $U_{1}(x,z)$ is decomposed into three
components:
\begin{eqnarray}
U_1(x,z)=U^i(x,z)+U^r(x,z)+U^d(x,z),
\end{eqnarray}
each of which satisfies the Helmholtz equation. $U^i$ represents the incident
field, which is assumed to be a plane wave of unit amplitude:
\begin{eqnarray}
U^i(x,z)=\exp(-ik_1z).
\end{eqnarray}
$U^r$ denotes the field that would be reflected if there were no slit in the screen
and thus satisfies
\begin{eqnarray}
U^r(x,z)=U^i(x,2b-z).
\end{eqnarray}
$U^d$ describes the diffracted field in region I due to the presence of the slit.
With the above set of equations and standard boundary conditions for a perfectly
conducting screen, a unique solution exists for the diffraction problem.
To find the field, the 2-dimensional Green's theorem is applied
with one function given by $U(x,z)$ and the other by a conventional Green's
function:
\begin{eqnarray}
({\nabla}^2+k_{j}^2)G_j=-{\delta}(x-x',z-z'),
\end{eqnarray}
where $(x,z)$ refers to a field point of interest; $x',z'$ are
integration variables, $j=1,2,3$. Since $U_j$ satisfies the
Helmholtz equation, Green's theorem reduces to
\begin{eqnarray}
U(x,z)=\int_{Boundary}(G{\partial}_n{U}-U{\partial}_n{G})dS.
\end{eqnarray}
The unknown fields $U^d(x,z)$, $U_3(x,z)$ and $U_2(x,z)$ are found
using the reduced Green's theorem and boundary conditions on $G$
\begin{eqnarray}
U^d(x,z)=-{\frac{\epsilon_1}{\epsilon_2}}\int_{-a}^{a}G_1(x,z;x',b)DU_b(x')dx'
\end{eqnarray}
for $b<z<\infty$,
\begin{eqnarray}
U_3(x,z)={\frac{\epsilon_3}{\epsilon_2}}\int_{-a}^{a}G_3(x,z;x',0)DU_0(x')dx'
\end{eqnarray}
for $-{\infty}<z<0$,
\begin{eqnarray}
U_2(x,z)=-\int_{-a}^{a}[G_2(x,z;x',0)DU_0(x')-
U_0(x'){\partial}_{z'}G_2(x,z;x',z')|_{z\rightarrow{0^+}}]dx\nonumber\\
+\int_{-a}^{a}[G_2(x,z;x,b)DU_b(x')-
U_b(x'){\partial}_{z'}G_2(x,z;x',z')|_{z\rightarrow{b^-}}]dx
\end{eqnarray}
for $|x|<a$ and $0<z<b$. The boundary fields in Eqs. 11-13 are defined by
\begin{eqnarray}
U_0(x)=U_2(x,z)|_{z\rightarrow{0^+}},\\
DU_0(x)={\partial}_zU_2(x,z)|_{z\rightarrow{0^+}},\\
U_b(x)=U_2(x,z)|_{z\rightarrow{b^-}},\\
DU_b(x)={\partial}_zU_2(x,z)|_{z\rightarrow{b^-}}.
\end{eqnarray}
In regions I and III the two Green's functions in Eqs. 11 and 13 are given by
\begin{eqnarray}
G_1(x,z;x',z')={\frac{i}{4}}[H_0^{(1)}(k_1R)+H_0^{(1)}(k_1R')],\\
G_3(x,z;x',z')={\frac{i}{4}}[H_0^{(1)}(k_3R)+H_0^{(1)}(k_3R'')],
\end{eqnarray}
with
\begin{eqnarray}
R=[(x-x')^2+(z-z')^2]^{1/2},\\
R'=[(x-x')^2+(z+z'-2b)^2]^{1/2},\\
R''=[(x-x')^2+(z+z')^2]^{1/2},
\end{eqnarray}
where $H_0^{(1)}$ is the Hankel function.
In region II, the Green's function in Eq. 12 is given by
\begin{eqnarray}
G_2(x,z;x',z')={\frac{i}{4a{\gamma_0}}}{\exp}(i{\gamma}_0|z-z'|)+
{\frac{i}{2a}}{\sum}_{m=1}^{\infty}{\gamma}_m^{-1}\nonumber\\
{\times}{\cos[m{\pi}(x+a)/2a]}{\cos[m{\pi}(x'+a)/2a]}\nonumber\\
{\times}{\exp(i{\gamma}m|z-z'|)},
\end{eqnarray}
where $\gamma_m=[k_2^2-(m{\pi}/2a)^2]^{1/2}.$ The field can be
found at any point once the four unknown functions in Eqs. 14-17
have been determined. The functions are completely determined by
a set of the four integral equations:
\begin{eqnarray}
2U_b^i(x)-U_b(x)={\frac{\epsilon_1}{\epsilon_2}}\int_{-a}^{a}G_1(x,b;x',b)DU_b(x')dx',
\end{eqnarray}
\begin{eqnarray}
U_0(x)={\frac{\epsilon_3}{\epsilon_2}}\int_{-a}^{a}G_3(x,0;x',0)DU_0(x')dx',
\end{eqnarray}
\begin{eqnarray}
{\frac{1}{2}}U_b(x)=-\int_{-a}^{a}[G_2(x,b;x',0)DU_0(x')-
U_0(x'){\partial}_{z'}G_2(x,b;x',z')|_{z\rightarrow{0^+}}]dx'\nonumber\\
+\int_{-a}^{a}[G_2(x,b;x',b)DU_b(x')]dx',
\end{eqnarray}
\begin{eqnarray}
{\frac{1}{2}}U_0(x)=-\int_{-a}^{a}[G_2(x,0;x',b)DU_b(x')
-U_b(x'){\partial}_{z'}G_2(x,0;x',z')|_{z\rightarrow{b_-}}]dx'\nonumber\\
-\int_{-a}^{a}[G_2(x,0;x',0)DU_0(x')]dx',
\end{eqnarray}
where $|x|<a$, and
\begin{eqnarray}
U_b^i(x)=\exp(-ik_1b).
\end{eqnarray}
The coupled integral equations 24-27 for the four boundary functions are
solved numerically. The magnetic ${\vec{H}}(x,z,t)$ and electric
${\vec{E}}(x,z,t)$ fields of the diffracted wave in region III are
found by using Eq. 12. The fields are given by
\begin{eqnarray}
{\vec{H}}(x,z,t)=i{\frac{a}{N}}{\frac{\epsilon_3}{\epsilon_2}}{\sum_{j=1}^N}
H_0^{(1)}[k_{3}((x-x_j)^2+z^2)^{1/2}]\nonumber\\
{\times}(D{\vec{U}}_0)_{j}{\exp}(-i\omega{t}){\vec{e}}_y,
\end{eqnarray}
\begin{eqnarray}
E_{x}(x,z,t)=-{\frac{a}{N}}{\frac{(\epsilon_3)^{1/2}}{\epsilon_2}}{\sum_{j=1}^N}
{\frac{z}{((x-x_j)^2+z^2)^{1/2}}}H_1^{(1)}[k_{3}((x-x_j)^2+z^2)^{1/2}]\nonumber\\
{\times}(D{\vec{U}}_0)_{j}{\exp}(-i\omega{t}),
\end{eqnarray}
\begin{eqnarray}
E_{y}(x,z,t)=0,
\end{eqnarray}
\begin{eqnarray}
E_{z}(x,z,t)={\frac{a}{N}}{\frac{(\epsilon_3)^{1/2}}{\epsilon_2}}{\sum_{j=1}^N}
{\frac{x-x_j}{((x-x_j)^2+z^2)^{1/2}}}H_1^{(1)}[k_{3}((x-x_j)^2+z^2)^{1/2}]\nonumber\\
{\times}(D{\vec{U}}_0)_{j}{\exp}(-i\omega{t}),
\end{eqnarray}
where $x_{j}=2a(j-1/2)/N-a$, $j=1,2,...,N$; $N>2a/z$; $H_1^{(1)}$
is the Hankel function. The coefficients $(D{\vec{U}}_0)_{j}$ are
found by solving numerically the four integral equations 24-27.
For more details of the model and the numerical solution of the
coupled integral equations 24-27 see refs. \cite{Neer,Betz1}.

Let us now consider the diffraction of an ultra-short pulse (wave
packet). The magnetic field of the incident pulse is assumed to be
Gaussian-shaped in time and both polarized and constant in the $y$
direction:
\begin{eqnarray}
{\vec{H}}(x,y,z,t)=U(x,z){\exp[-2\ln(2)(t/{\tau})^2]}{\exp}(-i\omega_{0}{t})
{\vec{e}}_y,
\end{eqnarray}
where $\tau$ is the pulse duration and
$\omega_0=2{\pi}c/\lambda_{0}$ is the central frequency. The pulse
can be composed in the wave-packet form of a Fourier time
expansion (for example, see ref. \cite{Kuk1,Kuk2}):
\begin{eqnarray}
{\vec{H}}(x,y,z,t)=\int_{-{\infty}}^{{\infty}}{\vec{H}}(x,z,\omega)
{\exp}(-i\omega{t})d\omega.
\end{eqnarray}
The electric and magnetic fields of the diffracted pulse are found
by using the expressions (1-32) for each  $\omega$-Fourier
component of the wave-packet (34). This algorithm is implemented
numerically by using the discrete fast fourier transform (FFT)
instead of the integral (34). The spectral interval
$[\omega_{min},\omega_{max}]$ and the sampling points $\omega_i$
are optimized by matching the FFT result to the original function
(33).
\section{Numerical analysis and discussion}
In this section, we test weather a light pulse can be resonantly
enhanced and simultaneously localized in space and time by a
subwavelength nano-sized metal slit. To address this question, the
spatial distribution of the energy flux of the transmitted pulse
under various regimes of the near-field diffraction is analyzed
numerically. The electric ${\vec{E}}$ and magnetic ${\vec{H}}$
fields of the transmitted pulse in the near-field diffraction zone
are computed by solving the equations (1-32) for each Fourier
$\omega$-component of the wave-packet (34). The amplitude of a FFT
$\omega$-component of the wave-packet transmitted through the slit
depends on the wavelength $\lambda=2{\pi}c/{\omega}$. Owing to the
dispersion, the Fourier spectra of the transmitted wave-packet
changes leading to modification of the pulse width and duration.
The dispersion of a continuous wave is usually described by the
normalized transmission coefficient $T_{cw}(\lambda)$, which is
calculated by integrating the normalized energy flux $S_z/S_z^i$
over the slit value~\cite{Neer,Betz1}:
\begin{eqnarray}
T_{cw}=-\frac{\sqrt{\epsilon_1}}{4a \cos{\theta}}\int_{-a}^{a}
{\lim_{z\rightarrow{0^-}}}[(E_{x}H_{y}^*+E_{x}^*H_{y})]dx,
\end{eqnarray}
where $S_z^i$ is the energy flux of the incident wave of unit
amplitude; $S_z$ is the transmitted flux. In order to establish
guidelines for the results of our numerical analysis, we computed
the transmission coefficient $T_{cw}({\lambda},a,b)$ for a
continuous wave ($\omega$-Fourier component) as a function of
screen thickness $b$ and/or wavelength $\lambda$ for different
values of slit width $2a$. Throughout the computations, the
magnitude of the incident magnetic field was assumed to be equal
to 1. As an example, the dependence $T_{cw}=T_{cw}(b)$ computed
for the wavelength $\lambda$=800 nm and the slit width $2a$ = 25
nm is shown in Fig. 2.  The dispersion $T_{cw}=T_{cw}(\lambda)$
for $2a$ = 25 nm and different values of the screen thickness $b$
is presented in Fig. 3. In Fig. 2, we note the transmission
resonances of $\lambda$/2 periodicity with the peak heights
$T_{cw}{\approx}{\lambda}/{2\pi}a$ at the resonances. Notice, that
the resonance positions and the peak heights are in agreement with
the results \cite{Harr,Betz1}. The dispersion
$T_{cw}=T_{cw}(\lambda)$ (curves A and B) presented in Fig. 3
indicates the wave-slit interaction behavior, which is similar to
those of a Fabry-Perot resonator. The transmission resonance
peaks, however, have a systematic shift towards longer
wavelengths. Our computations showed that the peak heights at the
main (strongest) resonant wavelength $\lambda_0^R$ (in the case of
Fig. 3, $\lambda_0^R$= 500 or 800 nm) are given by
$T_{cw}(\lambda_0^R,a){\approx}{\lambda_0^R}/{2\pi}a$. This
dependence indicates that the optical transmission of a
time-harmonic continuous wave can be enhanced by several orders of
magnitude by decreasing the slit width and increasing the
wavelength. Notice, that the Fabry-Perot like behaviour of the
transmission coefficient is in agreement with analytical and
experimental results published earlier \cite{Taka,Yang}.

The existence of transmission resonances for Fourier's
$\omega$-components of a wave-packet leads to the question: What
effect the resonant enhancement has on the spatial and temporal
localization of a light pulse? Presumably, the high transmission
at resonance occurs when the system efficiently channels Fourier's
components of the wave-packet from a wide area through the slit.
At resonance, one might assume that if the energy flow is
symmetric about the screen, the pulse width and duration should
increase very rapidly past the screen. Thus the large pulse
strength associated with resonance could only be obtained at the
expense of the spatial and temporal broadening of the wave-packet.
To test this hypothesis, the spatial distributions of the energy
flux of a transmitted wave-packet were computed for different slit
thickness corresponding to the resonance and anti-resonance
position.  As an example, Figs. 4 (a) and 5 (a) show the energy
flux of the anti-resonantly transmitted pulses. Figures 4 (b) and
5 (b) correspond to the case of the waveguide-mode resonance in
the slit. Figures 4 and 5 show the transmitted pulses at the
distances $|z|$ = $a/2$ and $a$, respectively. The comparison of
the flux distribution presented in Fig. 4 (a) with those of Fig.
4(b) shows that, for the parameter values adopted, a transmitted
wavepacket is enhanced by one order of magnitude and
simultaneously localized in the 25-nm and 100-fs domains of the
near-field diffraction zone. Thus at the distance $|z|$ = $a/2$,
the slit resonantly enhances the intensity of the pulse without
its spatial and temporal broadening. The result can be easily
understood by considering the dispersion properties of the slit.
For the screen thickness $b$ = 200 nm, the amplitudes of Fourier's
components of the wave-packet, whose central wavelength
$\lambda_0$ is detuned from the main (at 500 nm) resonance, are
practically unchanged in the wavelength region near 800 nm (see,
curve B in Fig. 3). This provides the dispersion- and
distortion-free non-resonant transmission of the wave-packet
(Figs. 4(a) and 5(a)). In the case of the thicker screen (b = 350
nm), the slit transmission experiences strong mode-coupling regime
at the wavelengths near to 800 nm (see, curve A of Fig. 3) that
leads to a profound and uniform enhancement of amplitudes all of
the Fourier $\omega$-components of the wave-packet (see curve C in
Fig. 3). Thus, the slit resonantly enhances by one order of
magnitude the intensity of the pulse without its spatial and
temporal broadenings (Figs. 4(a) and 5(a)). Also, notice that at
the distance $|z|$ = $a$ (Figs. 5(a) and 5(b)), both the
resonantly and anti-resonantly transmitted pulses experience
natural spatial broadening in the transverse direction, while
their durations are practically unchanged.

By comparing the data for anti-resonant and resonant transmissions
presented in Figs. 4 and 5 one can see that at the appropriate
values of the distance $|z|$ and the wave-packet spectral width
$\Delta{\omega}$ the resonance effect does not influence the
spatial and temporal localization of the wave-packet. To verify
this somewhat unexpected result, the FWHMs of the transmitted
pulse in the transverse and longitudinal directions were
calculated for different values of the slit width $2a$, central
wavelength $\lambda_0$ and pulse duration
${\tau}{\approx}1/\Delta{\omega}_p$ as a function of screen
thickness $b$ at two particular near-field distances $|z|$ = $a/2$
and $a$ from the screen. It was seen that, at the dispersion-free
resonant transmission condition $\Delta{\omega}_p
<0.1\Delta{\omega}_r$ (here, $\Delta{\omega_r}$ is the resonance
spectral width), the transmitted pulse indeed does not experience
temporal broadening. Thus the temporal localization associated
with the duration $\tau$ of the incident pulse remains practically
unchanged under the transmission. The value of $\tau$ is
determined by the dispersion-free condition
${\tau}{\approx}1/\Delta{\omega}_p>1/0.1\Delta{\omega}_r$, where
$\Delta{\omega}_r=\Delta{\omega}_r(a)$ practically does not depend
on the screen thickness $b$.  We found that the energy flux of the
transmitted wave-packet can be enhanced by a factor
$T_{cw}(\lambda_0^R,a){\approx}{\lambda_0^R}/{2\pi}a$ by the
appropriate adjusting the screen thickness $b=b(\lambda_0^R)$, for
an example see Figs. 3-5. Thus the wave-packet can be enhanced by
a factor ${\lambda_0^R}/{2\pi}a$ and simultaneously localized in
the time domain at the ${\tau}={\tau}(a)$ scale. It was also seen
that the FWHM of the transmitted pulse in the transverse direction
depends on the wave-packet central wavelength $\lambda_0$ and the
distance $z$ from the slit. Nevertheless, the FWHM of the
transmitted pulse can be always reduced to the value $2a$ by the
appropriate decreasing the distance $|z|$ = $|z|(a)$ from the
screen ($|z|$ = $a/2$, in the case of Fig. 4). Thus high
transmission can be achieved without concurrent loss in the degree
of temporal and spatial localizations of the pulse. In retrospect,
this result is reasonable, since the symmetry of the problem for a
time-harmonic continuous wave (Fourier's $\omega$-component of a
wave-packet) is disrupted by the presence of the initial and
reflected fields in addition to the diffracted field on one side
(Eq. 6). As the thickness changes, the field $U^d$ and $U_3$
change only in magnitude, but the field $U_1$ changes in
distribution as well since it involves the sum of $U^d$ with
unchanging fields $U^i$ and $U^r$. At resonance, the distribution
of $U_1$ leads to channeling of the radiation, but the
distribution of $U_3$ remains unaffected. By the appropriate
adjusting the slit-pulse parameters a light can be enhanced by
orders of magnitude and simultaneously localized in the near-field
diffraction zone at the nm- and fs-scales.

The limitations of the above analysis must be considered before
the results are used for a particular experimental device. The
resonant enhancement with simultaneous nm-scale spatial and
fs-scale temporal localizations of a light by a subwavelength
metal nano-slit is a consequence of the assumption of the screen
perfect conductivity. The slit can be made of perfectly conductive
(at low temperatures) materials. In the context of current
technology, however, the use of conventional materials like metal
films at a room temperature is more practical. As a general
criterion, the perfect conductivity assumption should remain valid
as far as the slit width and the screen thickness exceed the
extinction length for the Fourier $\omega$-components of a
wavepacket within the metal. The light intensity decays in the
metal screen at the rate of $I_s = I_0 {\exp}(-b/{\delta})$, where
${\delta}$ = ${\delta}(\lambda)$ is the extinction length in the
screen. The aluminum has the largest opacity ($\delta<$ 11 nm) in
the spectral region $\lambda >$ 100 nm [20]. The extinction length
increases from 11 to 220 nm with decreasing the wavelength from
100 to 50 nm. Hence, the perfect conductivity is a very good
approximation in a situation involving a relatively thick ($b >$
25 nm) aluminum screen and a wave-packet of the duration
${\tau}{\approx}1/\Delta{\omega}_p$ having the Fourier components
in the spectral region $\lambda >$ 100 nm. However, in the case of
thinner screens, shorter pulses and smaller central wavelengths of
wave-packets, the metal films are not completely opaque. This
would reduce a value of spatial localization of a pulse due to
passage of the light through the screen in the region away from
the slit. Moreover, the phase shifts of the Fourier components
along the propagation path causing by the skin effect can modify
the enhancement coefficient and temporal localization properties
of the slit.

The above analysis is directly applicable to the two-dimensional
near-field scanning optical microscopy and spectroscopy. In a
conventional 2D NSOM, a subwavelength ($2a<\lambda$) slit
illuminated by a continuous wave is used as a near-feld ($|z|<a$)
light source providing the nm-scale resolution in space
\cite{Ash,Lewi,Betz2,Pohl}. The non-resonant transmission of fs
pulses could provide the super resolution of NSOM simultaneously
in space and time \cite{Kuk1,Kuk2}. The above-described resonantly
enhanced transmission together with nm- and fs-scale localizations
in the space and time of a pulse could greatly increase a
potential of the near-field scanning optical microscopy and
spectroscopy, especially in high-resolution applications. It
should be noted in this connection that the high transmission
($T_{cw}(\lambda_0^R,a){\approx}{\lambda_0^R}/{2\pi}a$) of a pulse
can be achieved without concurrent loss in the temporal and
spatial localizations of the pulse only at the short
($|z|=|z|(a)$) distances from the slit. The presence of a
microscopic sample (a molecule, for example) placed at the short
distance in strong interaction with NSOM slit, however, modifies
the boundary conditions. In the case of the strong
slit-sample-pulse interaction, which takes place at the distance
$|z|<<0.1a$, the response function accounting for the modification
of the quantum mechanical behavior of the sample should be took
into consideration. The potential applications of the effect of
the resonantly enhanced transmission together with nm-scale and
fs-scale localizations of a pulse are not limited to near-field
microscopy and spectroscopy. Broadly speaking, the effect concerns
all physical phenomena and photonic applications involving a
transmission of light by a single subwavelength nano-slit, a
grating with subwavevelength slits and a subwavelength slit
surrounded by parallel grooves (see, the
studies~\cite{Neer,Harr,Betz1,Ash,Lewi,Betz2,Pohl,Ebbe,Leze,Port,Hibb,
Gar1,Gar2,Dog,Li,Stoc,Kuk1,Kuk2,Stav,Bori,Taka,Yang,Dykh,Cao,Alte}
and references therein). For instance, the effect could be used
for sensors, communications, optical switching devices and
microscopes.

\section{Conclusion}
In conclusion, in the present article we have considered a
question weather a light can be enhanced and simultaneously
localized in space and time by a subwavelength metal nano-slit. To
address this question, the spatial distributions of the energy
flux of an ultrashort (fs) pulse diffracted by a subwavelength
(nanosized) slit in a thick metal screen of perfect conductivity
have been analyzed by using the conventional approach based on the
Neerhoff and Mur solution of Maxwell's equations. The analysis of
the spatial distributions for various regimes of the near-field
diffraction demonstrated that the energy flux of a wavepacket can
be enhanced by orders of magnitude and simultaneously localized in
the near-field diffraction zone at the nm- and fs-scales. The
extraordinary transmission, together with the nm-scale and
fs-scale localizations of a light, make the nano-slit structures
attractive for many photonic purposes, such as sensors,
communications, optical switching devices and NSOM. We also
believe that the addressing of the above-mentioned basic question
gains insight into the physics of near-field resonant diffraction.
\begin{acknowledgments}
This study was supported by the Fifth Framework of the European
Commission (Financial support from the EC for shared-cost RTD
actions: research and technological development projects,
demonstration projects and combined projects. Contract
NG6RD-CT-2001-00602). The authors thank the Computing Services
Centre, Faculty of Science, University of Pecs, for providing
computational resources.
\end{acknowledgments}
\newpage
\begin{figure}
\includegraphics[keepaspectratio,width=13cm]{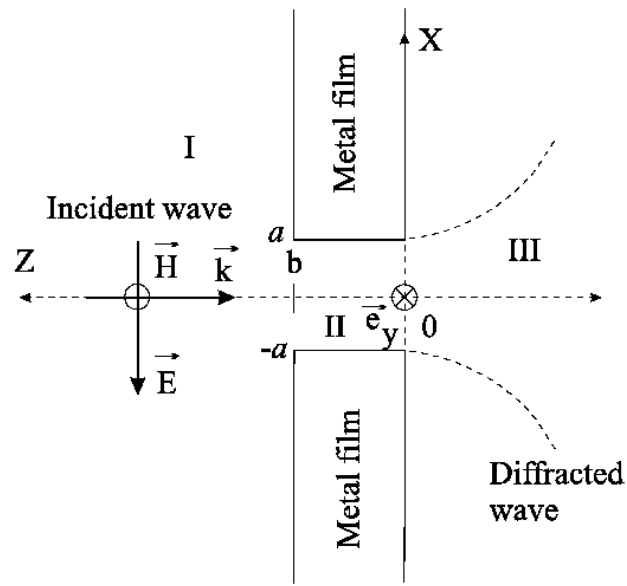}
\caption{\label{fig:epsart}Propagation of a continuous wave
through a subwavelength nano-sized slit in a thick metal film.}
\end{figure}
\newpage
\begin{figure}
\includegraphics[keepaspectratio,width=13cm]{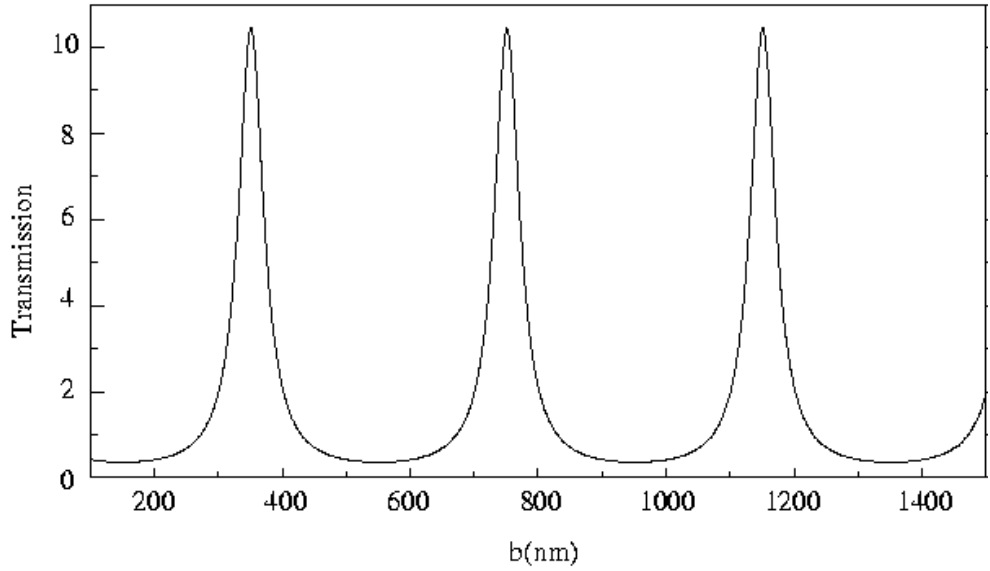}
\caption{\label{fig:epsart}The transmission coefficient $T_{cw}$
for a continuous wave ($\omega$-Fourier component of a
wave-packet) as a function of screen thickness $b$ computed for
the wavelength $\lambda$=800 nm and the slit width $2a$ = 25 nm.}

\end{figure}
\newpage
\begin{figure}
\includegraphics[keepaspectratio,width=13cm]{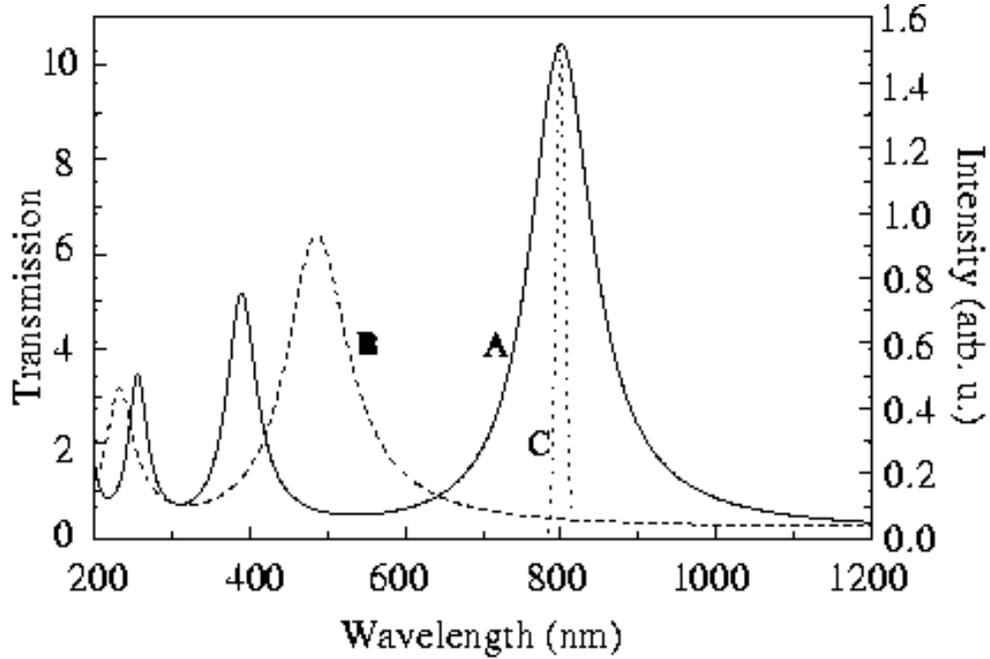}
\caption{\label{fig:epsart}The dispersion $T_{cw}=T_{cw}(\lambda)$
for a continuous wave ($\omega$-Fourier component of a
wave-packet) computed for the slit width $2a$ = 25 nm and
different values of the screen thickness $b$: A - 350 nm and B -
200 nm. The Fourier spectra (curve C) is presented for the
comparison. The curve C shows the Fourier spectra of an incident
wave-packet with the duration ${\tau}=100$ fs and the central
wavelength $\lambda_0=$ 800 nm, which was used in the computations
presented in Fig. 4 and 5.}
\end{figure}
\newpage
\begin{figure}
\includegraphics[keepaspectratio,width=13cm]{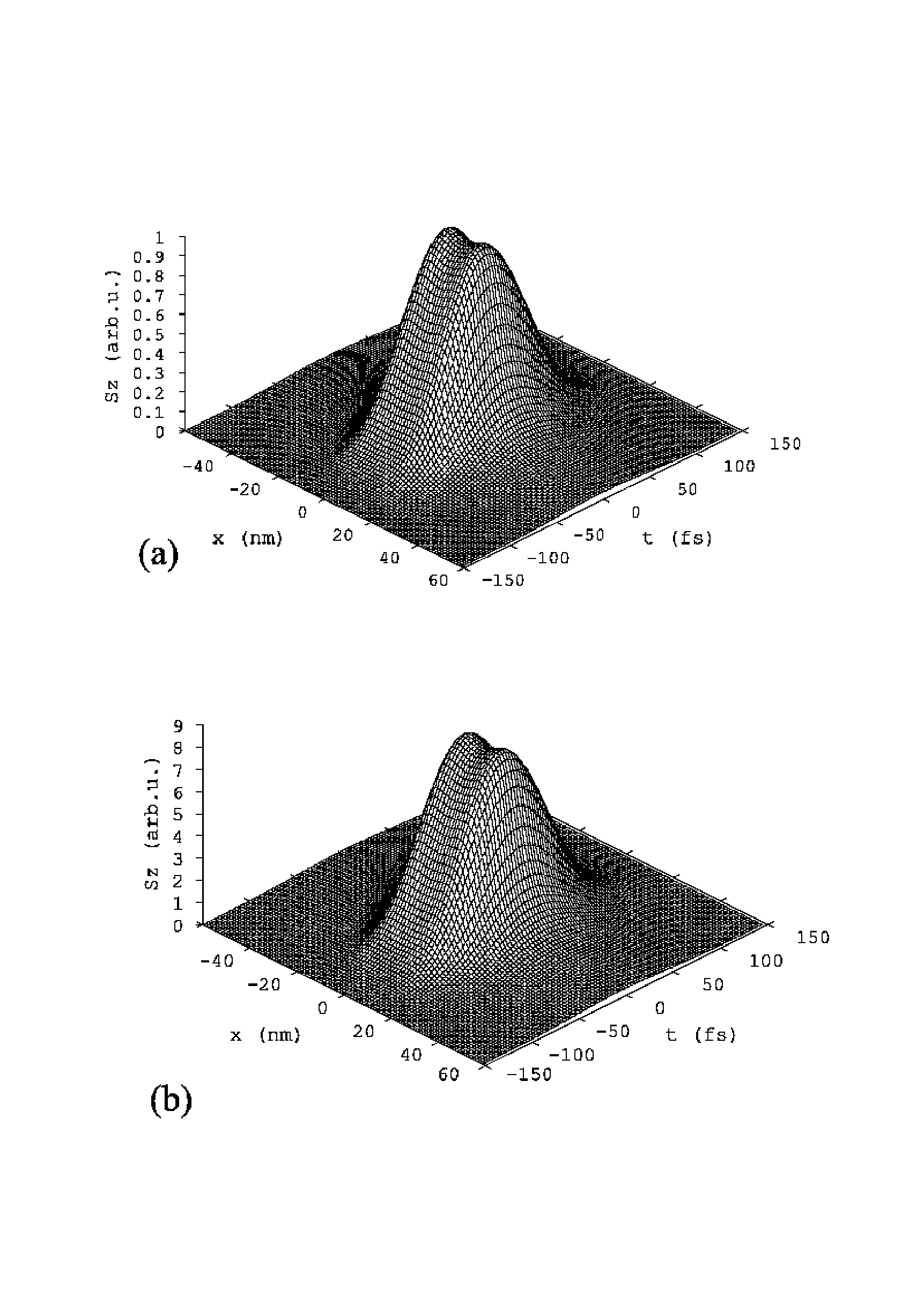}
\caption{\label{fig:epsart}The energy flux of a transmitted pulse
at the distances $|z|$ = $a/2$. (a) The anti-resonant transmission
by the slit ($2a$ = 25 nm, $b$ = 200 nm). (b) The resonant
transmission by the slit ($2a$ = 25 nm, $b$ = 350 nm). The
incident wave-packet duration ${\tau}=100$ fs and the central
wavelength $\lambda_0=$ 800 nm.}
\end{figure}
\newpage
\begin{figure}
\includegraphics[keepaspectratio,width=13cm]{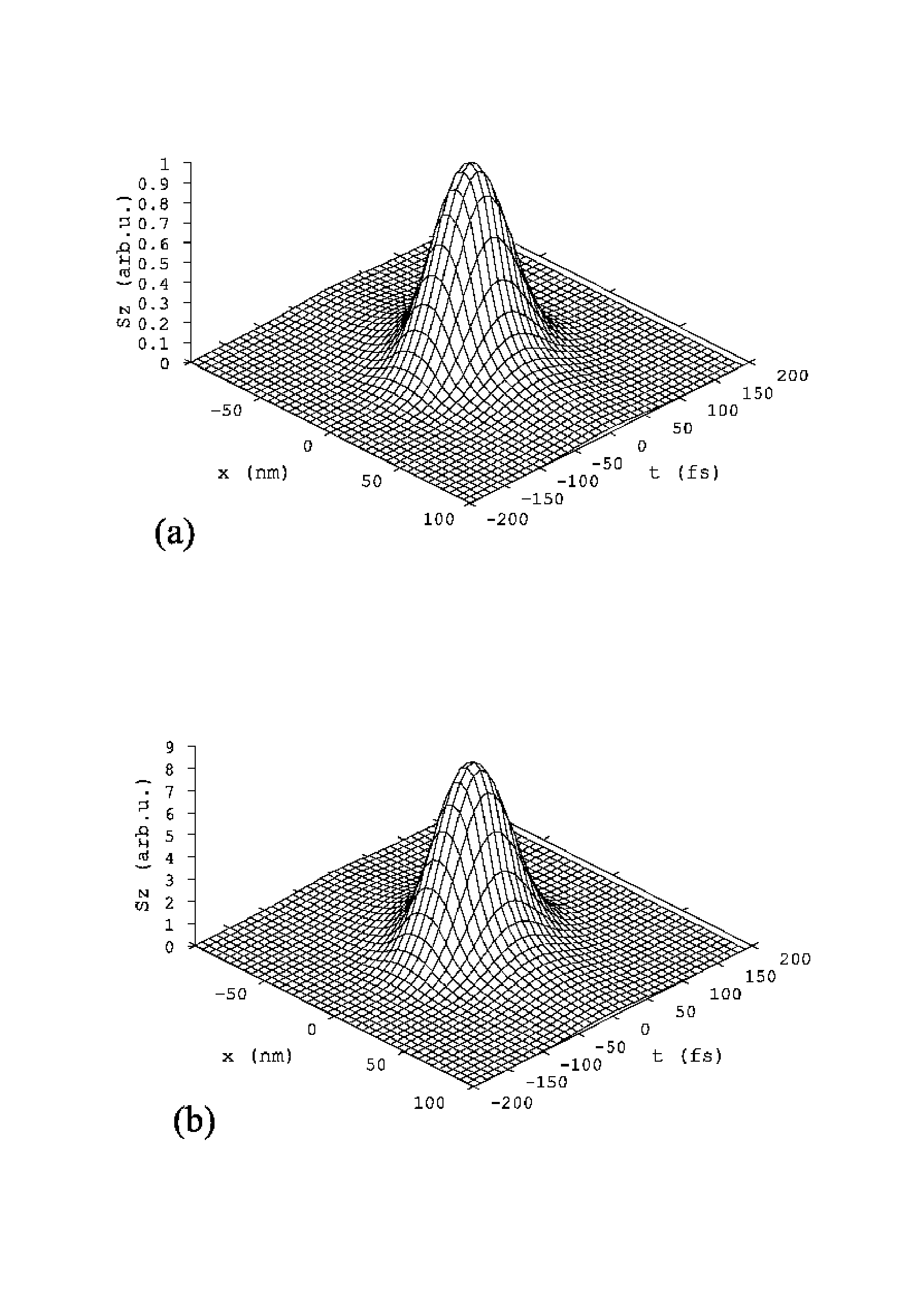}
\caption{\label{fig:epsart}The energy flux of a transmitted pulse
at the distances $|z|$ = $a$. (a) The anti-resonant transmission
by the slit ($2a$ = 25 nm, $b$ = 200 nm). (b) The resonant
transmission by the slit ($2a$ = 25 nm, $b$ = 350 nm). The
incident wave-packet duration ${\tau}=100$ fs and the central
wavelength $\lambda_0=$ 800 nm.}
\end{figure}
\end{document}